\documentclass[12pt]{article}
\usepackage{a4wide,epsfig,amsmath}

\newcommand{\agt}{\rlap{\lower 3.5 pt \hbox{$\mathchar \sim$}} \raise 1pt
 \hbox {$>$}}
\newcommand{\alt}{\rlap{\lower 3.5 pt \hbox{$\mathchar \sim$}} \raise 1pt
 \hbox {$<$}}
\newcommand{\re}{\mathop{\mathrm{Re}}\nolimits}
\newcommand{\im}{\mathop{\mathrm{Im}}\nolimits}

\catcode`@=11
\newcount\@tempcntc
\def\@citex[#1]#2{\if@filesw\immediate\write\@auxout{\string\citation{#2}}\fi
  \@tempcnta\z@\@tempcntb\m@ne\def\@citea{}\@cite{\@for\@citeb:=#2\do
    {\@ifundefined
       {b@\@citeb}{\@citeo\@tempcntb\m@ne\@citea\def\@citea{,}{\bf
?}\@warning
       {Citation `\@citeb' on page \thepage \space undefined}}%
    {\setbox\z@\hbox{\global\@tempcntc0\csname b@\@citeb\endcsname\relax}%
     \ifnum\@tempcntc=\z@ \@citeo\@tempcntb\m@ne
       \@citea\def\@citea{,}\hbox{\csname b@\@citeb\endcsname}%
     \else
      \advance\@tempcntb\@ne
      \ifnum\@tempcntb=\@tempcntc
      \else\advance\@tempcntb\m@ne\@citeo
      \@tempcnta\@tempcntc\@tempcntb\@tempcntc\fi\fi}}\@citeo}{#1}}
\def\@citeo{\ifnum\@tempcnta>\@tempcntb\else\@citea\def\@citea{,}%
  \ifnum\@tempcnta=\@tempcntb\the\@tempcnta\else
   {\advance\@tempcnta\@ne\ifnum\@tempcnta=\@tempcntb \else
\def\@citea{--}\fi
    \advance\@tempcnta\m@ne\the\@tempcnta\@citea\the\@tempcntb}\fi\fi}
\catcode`@=12

\begin{document}

\title{
\vskip-3cm{\baselineskip14pt
\centerline{\normalsize DESY 12--119\hfill ISSN 0418-9833}
\centerline{\normalsize June 2012\hfill}}
\vskip1.5cm
Low-mass Higgs decays to four leptons at one loop and beyond}

\author{Bernd A. Kniehl and Oleg L. Veretin\\
{\normalsize II. Institut f\"ur Theoretische Physik, Universit\"at Hamburg,}\\
{\normalsize Luruper Chaussee 149, 22761 Hamburg, Germany}
}

\date{}

\maketitle

\begin{abstract}
The ongoing searches for Higgs-boson signals in data taken at the CERN LHC and
the Fermilab Tevatron crucially rely on the decay channels $H\to Z\ell\ell$ and
$H\to W\ell\nu_l$.
We present a precision study of the partial widths of these decay channels
including the full one-loop electroweak corrections and the dominant
contributions at two and three loops, of ${\cal O}(G_F^2m_t^4)$,
${\cal O}(G_Fm_t^2\alpha_s)$, and ${\cal O}(G_Fm_t^2\alpha_s^2)$.
Since the invariant mass of the off-shell intermediate boson is relatively low
in the mass window $115~\text{GeV}<m_H<129~\text{GeV}$ of current interest,
lepton mass effects are relevant, especially for the $\tau$ lepton. 
\medskip

\end{abstract}

\newpage

\section{Introduction}

The ATLAS and CMS collaborations at the $pp$ collider LHC are presently closing
in on the Higgs boson of the standard model (SM).
The combined ATLAS results, based on the full data set of up to 4.9~fb${}^{-1}$
recorded during the 2011 operation at the center-of-mass energy
$\sqrt{s}=7$~TeV, excluded the Higgs-boson mass ($m_H$) windows from
110.0~GeV to 117.5~GeV, from 118.5~GeV to 122.5~GeV, and from 129~GeV to
539~GeV at the 95\% confidence level (C.L.) \cite{ATLAS:2012ae}.
A small excess of signal events over the estimated background was observed
around $m_H=126$~GeV with a local significance of 2.5 standard deviations
($\sigma$).
The CMS Collaboration excluded the $m_H$ window from 127.5~GeV to 600~GeV at
95\% C.L. exploiting approximately 5~fb${}^{-1}$ of data at $\sqrt{s}=7$~TeV
\cite{Chatrchyan:2012tx}.
They observed some excess in the $m_H$ window from 115~GeV to 128~GeV, with a
maximum local significance of 2.8~$\sigma$ at $m_H=125$~GeV.

The CDF and D0 collaborations at the $p\overline{p}$ collider Tevatron jointly
excluded the $m_H$ windows from 100~GeV to 106~GeV and from 141~GeV to 184~GeV
at 95\% C.L. by analyzing a luminosity of up to 10~fb${}^{-1}$ collected at
$\sqrt{s}=1.96$~TeV \cite{TEVNPH:2012ab}.
They observed an excess of signal over background in the $m_H$ window from
115~GeV to 135~GeV, with a local significance of 2.2~$\sigma$ at $m_H=120$~GeV.
The ALEPH, DELPHI, L3, and OPAL collaborations at the CERN $e^+e^-$ collider
LEP established a $m_H$ lower bound of 114.4~GeV at 95\% C.L. using a total
of 2.461~fb${}^{-1}$ data collected at 189~GeV${}<\sqrt{s}<209$~GeV
\cite{Barate:2003sz}.   

The searches at the LHC for the SM Higgs boson in the low-$m_H$ region mainly
rely on its decay channels
$H\to\gamma\gamma$ \cite{Aad:2011ww,Chatrchyan:2012tw},
$H\to ZZ^\ast\to\ell^+\ell^-\ell^{\prime+}\ell^{\prime-}$
\cite{Aad:2011uq,Chatrchyan:2012dg},
$H\to W^+W^{-\ast},W^{+\ast}W^-\to\ell^+\nu_\ell\ell^{\prime-}
\overline{\nu}_{\ell^\prime}$
\cite{ATLAS:2011aa,Chatrchyan:2011tz},
where off-mass-shell particles are marked by asterisks and
$\ell,\ell^\prime=e,\mu$.
Decay channels of lesser significance include $H\to\tau^+\tau^-$
\cite{Chatrchyan:2012vp} and, in $W^\pm H$ and $ZH$ associated production with
subsequent
$W^\pm\to\ell^\pm{}_{\vphantom\ell}^{\scriptscriptstyle(}%
\overline{\nu}_\ell^{\scriptscriptstyle)}$
and $Z\to\ell^+\ell^-$ decays, $H\to b\overline{b}$ \cite{Chatrchyan:2012ww}.
In the following, we briefly review the state of the art regarding the
theoretical predictions for the respective partial decay widths.
Comprehensive reviews of our theoretical knowledge of the SM Higgs boson may
be found in Refs.~\cite{Kniehl:1993ay,Spira:1997dg} and references cited
therein.

As for the partial decay width $\Gamma(H\to\gamma\gamma)$, the lowest-order
result was first obtained in Ref.~\cite{Ellis:1975ap}.
The two-loop ${\mathcal O}(\alpha_s)$ \cite{Zheng:1990qa} and three-loop
${\mathcal O}(\alpha_s^2)$ \cite{Steinhauser:1996wy} QCD corrections are
available.
As for the two-loop ${\mathcal O}(\alpha)$ correction, the contributions
induced by light \cite{Aglietti:2004nj} and heavy fermions
\cite{Djouadi:1997rj,Fugel:2004ug} as well as the residual ones
\cite{Degrassi:2005mc} are known.

As for the partial decay width $\Gamma(H\to b\overline{b})$, the one-loop
${\mathcal O}(\alpha_s)$ \cite{Braaten:1980yq} and ${\mathcal O}(\alpha)$
\cite{Fleischer:1980ub,Bardin:1990zj} corrections have been known for a long
time.
As for the two-loop ${\cal O}(\alpha_s^2)$ correction, the leading
\cite{Gorishnii:1990zu} and next-to-leading \cite{Surguladze:1994gc} terms of
the expansion in $m_b^2/m_H^2$ of the diagrams without top quarks are known.
The diagrams containing a top quark can be divided into two classes:
diagrams containing gluon self-energy insertions, which are exactly known
\cite{Kniehl:1994vq}, and double-triangle diagrams, for which the four leading
terms of the expansion in $m_H^2/m_t^2$ are known \cite{Chetyrkin:1995pd}. 
The $m_b$-independent terms of the two-loop ${\cal O}(\alpha^2)$ and
${\cal O}(\alpha\alpha_s)$ corrections may be obtained from the analogous
${\cal O}(\alpha_s^2)$ one \cite{Gorishnii:1990zu} by adjusting coupling
constants and color factors \cite{Kataev:1997cq}.
As for the two-loop correction of order ${\cal O}(x_t\alpha_s)$, where
$x_t=G_Fm_t^2/(8\pi^2\sqrt{2})$ parametrizes leading power corrections of
top-quark origin, the universal part, which appears for any Higgs-boson decay
to a fermion pair, was extracted from the full ${\cal O}(\alpha\alpha_s)$
result \cite{Kniehl:1993jc} and the nonuniversal one, which arises because
bottom is the isopartner of top, using a low-energy theorem
\cite{Kwiatkowski:1994cu}.
As for the two-loop ${\cal O}(x_t^2)$ correction, the universal
\cite{Djouadi:1997rj,Butenschoen:2006ns} and nonuniversal
\cite{Butenschoen:2006ns} parts are both available.
The ${\cal O}(G_F^2m_H^4)$ term \cite{Durand:1994pk}, which is universal,
dominates the electroweak two-loop correction in the large-$m_H$ regime.
The three-loop ${\cal O}(\alpha_s^3)$ correction without top-quark
contribution was calculated in the massless limit \cite{Chetyrkin:1996sr}.
The correction induced by the top quark was subsequently found using an
appropriate effective field theory \cite{Chetyrkin:1997vj}.
As for the three-loop ${\cal O}(x_t\alpha_s^2)$ correction, the universal
\cite{Kniehl:1995at} and nonuniversal \cite{Chetyrkin:1996wr} parts are both
known.
The four-loop ${\cal O}(\alpha_s^4)$ correction without top-quark contribution
was calculated in the massless limit \cite{Baikov:2005rw}.
The residual theoretical uncertainty in the prediction of
$\Gamma(H\to b\overline{b})$ was assessed in Ref.~\cite{Kataev:2009ns}.

As for the partial decay width $\Gamma(H\to\tau^+\tau^-)$, we know the full
one-loop ${\mathcal O}(\alpha)$ \cite{Fleischer:1980ub,Bardin:1990zj} and
two-loop ${\mathcal O}(\alpha\alpha_s)$ \cite{Kniehl:1993jc} corrections as
well as the dominant terms of ${\mathcal O}(x_t^2)$ \cite{Djouadi:1997rj} and
${\mathcal O}(G_F^2m_H^4)$ \cite{Durand:1994pk} at two loops, and the one of
${\mathcal O}(x_t\alpha_s^2)$ \cite{Kniehl:1995at} at three loops.

The decays of the Higgs boson to two pairs of light fermions proceed almost
exclusively via $W^{+(\ast)}W^{-(\ast)}$ or $Z^{(\ast)}Z^{(\ast)}$ pairs, the
contributions involving the Yukawa couplings of the produced fermions being
greatly suppressed, as will be explicitly demonstrated for $H\to Z\tau^+\tau^-$
and $H\to W^+\tau^-\overline{\nu}_\tau$ in Sec.~\ref{sec:three}.
If the value of $m_H$ is large enough to allow for one or both intermediate
bosons to be on mass shell, then such kinematic configurations will be
extremely favored due to the resonating propagators \cite{Kniehl:1990yb}.
In this case, it is natural to employ the narrow-width approximation, which
implies that the resonating intermediate bosons are treated as real particles
and their subsequent decays are accounted for by multiplication with the
appropriate branching fractions.
Such a procedure is also routinely employed when the cross section of a
complete scattering process involving the unstable Higgs boson is factorized
into the cross section of the Higgs-boson production mechanism and the
branching fraction of the Higgs-boson decay channel.
In the case of Higgs-boson decays to two light-fermion pairs, this has the
advantage that the branching fractions $B(Z\to f\overline{f})$ and
$B(W^\pm\to f\overline{f^\prime})$ may be taken to be the experimentally measured
values, which naturally contain all radiative corrections.
As for the partial decay widths $\Gamma(H\to W^+W^-)$ and $\Gamma(H\to ZZ)$,
which presuppose that $m_H>2m_V$ with $V=W,Z$, the full one-loop
${\mathcal O}(\alpha)$ corrections
\cite{Fleischer:1980ub,Kniehl:1991xe,Kniehl:1990mq,Bardin:1991dp}, also
including the subsequent decays into massless-fermion pairs for off-shell
$V$ bosons \cite{Bredenstein:2006rh}, the dominant two-loop terms of
${\mathcal O}(x_t\alpha_s)$ \cite{Kniehl:1995ra,Kniehl:1995gj},
${\mathcal O}(x_t^2)$ \cite{Djouadi:1997rj}, and
${\mathcal O}(G_F^2m_H^4)$ \cite{Ghinculov:1995bz}, and the three-loop term of
${\mathcal O}(x_t\alpha_s^2)$ \cite{Kniehl:1995at} are available.
In Ref.~\cite{Bredenstein:2006rh}, the ${\mathcal O}(G_F^2m_H^4)$
corrections \cite{Ghinculov:1995bz}, which are only relevant for $m_H\gg2m_V$,
and certain higher-order effects due to photonic final-state radiation off
charged leptons, which are not logarithmically enhanced for the integrated
partial decay widths, were also included.

The purpose of this paper is to present a precision study of the partial decay
widths $\Gamma(H\to W^\pm f\overline{f^\prime})$ and
$\Gamma(H\to Zf\overline{f})$, appropriate for the mass window $m_V<m_H<2m_V$
of topical interest, retaining the masses of the produced fermions and
including the full one-loop corrections and the dominant higher-order terms.
In the case of $H\to Zf\overline{f}$ with massless final-state fermions, the
one-loop weak correction may be obtained \cite{Kniehl:1993ay} by crossing
symmetry from the corresponding analysis of $e^+e^-\to ZH$
\cite{Kniehl:1991hk,Denner:1992bc}.
Here, we redo this calculation for massive final-state fermions and incorporate
the ${\mathcal O}(x_t^2)$, ${\mathcal O}(x_t\alpha_s)$, and
${\mathcal O}(x_t\alpha_s^2)$ terms.
We also perform the analogous analysis for $H\to  W^\pm f\overline{f^\prime}$.
In fact, the numerical analyses for $H\to Z\tau^+\tau^-$ and
$H\to W^+\tau^-\overline{\nu}_\tau$ in Sec.~\ref{sec:three} will reveal that
the finite $\tau$-lepton mass effects may exceed the radiative corrections in
size in the low-$m_H$ window that is not excluded experimentally.
Both channels are bound to be exploited for the search of a low-$m_H$ Higgs
boson and/or the study of its properties in the long run.
The $H\to\ell^+\ell^-\tau^+\tau^-$ channel is already being used by the CMS
Collaboration in the high-$m_H$ range \cite{Chatrchyan:2012hr}.

This paper is organized as follows.
In Sec.~\ref{sec:two}, we expose the structure of our analytical results and
list an appropriate selection of our formulas.
In Sec.~\ref{sec:three}, we present our numerical analysis for the
$H\to Z\ell^+\ell^-$ and $H\to W^+\ell^-\overline{\nu}_\ell$ decay channels,
which are relevant for the ongoing searches for Higgs-boson signals in data
taken at the LHC and the Tevatron.
In Sec.~\ref{sec:four}, we summarize our conclusions.

\section{Analytic results}
\label{sec:two}

We now present our analytic results.
We work in the on-mass-shell renormalization scheme implemented with Fermi's
constant $G_F$, instead of Sommerfeld's fine-structure constant $\alpha$, and
use the shorthand notations $h=m_H^2$, $w=m_W^2$, $z=m_Z^2$, and
$c_w^2=1-s_w^2=w/z$.

\boldmath
\subsection{$H\to Zf\overline{f}$ decay}
\label{sec:twoa}
\unboldmath

We first consider the decay process $H\to Zf\overline{f}$ for a generic fermion
$f$ of mass $m$, electric charge $Q$, and color multiplicity $N$, which is
$N=1$ for leptons and $N=3$ for quarks.
The $Zf\overline{f}$ vector and axial-vector couplings are proportional to
$V=2I-4s_w^2Q$ and $A=2I$, where $I=\pm1/2$ is the third component of weak
isospin of the left-handed component of $f$.
Formally, the partial decay width of $H\to Zf\overline{f}$ and the one of
$Z\to Hf\overline{f}$, the radiative corrections to which were studied in
Ref.~\cite{Kniehl:1992qq}, are related as \cite{Kniehl:1993ay}
\begin{equation}
\Gamma_{H\to Zf\overline{f}}=-3\left(\frac{z}{h}\right)^{3/2}
\Gamma_{Z\to Hf\overline{f}},
\label{eq:rel}
\end{equation}
where the minus sign ensures that the phase space remains positive upon the
interchange $z\leftrightarrow h$, and the factors 3, $\sqrt{z/h}$, and $z/h$  
adjust the spin average, flux, and phase space, respectively.
In contrast to Ref.~\cite{Kniehl:1992qq}, we include here also the finite-$m$
corrections.

The distribution of the partial decay width $\Gamma_{H\to Zf\overline{f}}$ in
the $f\overline{f}$ invariant mass square $s=q^2$ may be written as
\begin{equation}
\frac{d\Gamma_{H\to Zf\overline{f}}}{ds}=
\frac{d\Gamma_{H\to Zf\overline{f}}^0}{ds}\left[\delta_0
+\frac{3}{4}\left(\frac{\alpha}{\pi}Q^2+\frac{\alpha_s}{\pi}C_F\right)\delta_1
+\delta_\text{w}-\delta_{x_t}+\delta_\text{res}\right]\delta_t,
\label{eq:dg}
\end{equation}
where
\begin{equation}
\frac{d\Gamma_{H\to Zf\overline{f}}^0}{ds}=\frac{G_F^2z^3}{128\pi^3h^{3/2}}
N(V^2+A^2)\frac{C_1}{(s-z)^2},
\label{eq:dg0}
\end{equation}
with $C_1=\sqrt{\lambda}[4s+\lambda/(3z)]$ \cite{Kniehl:1991hk} and
$\lambda=s^2+z^2+h^2-2(sz+zh+hs)$, is the tree-level result for $m=0$
\cite{Pocsik:1980ta,Keung:1984hn},
the factor $\delta_0$ restores the full $m$ dependence of the latter,
$C_F=(N^2-1)/(2N)$,
$\delta_1$ is the coefficient shared by the ${\cal O}(\alpha)$ QED and
${\cal O}(\alpha_s)$ QCD corrections, $\delta_\text{w}$ contains the purely
weak correction at one loop, $\delta_{x_t}$ is the leading ${\cal O}(x_t)$ term
of the latter, the factor $\delta_t$ supplies the leading top-quark-induced
corrections at one loop and beyond, and $\delta_\text{res}$ comprises the
residual higher-order corrections, which are beyond the scope of our present
analysis.

It is useful to distinguish between the class of contributions devoid of the
$Hf\overline{f}$ Yukawa coupling and the complementary class.
The former survives in the massless limit $m\to0$, and the parts of $\delta_0$
and $\delta_1$ that belong to it are related via the optical theorem to the
absorptive part of the $Z$-boson vacuum polarization tensor induced by the
fermion $f$,
\begin{equation}
\Pi^{\mu\nu}(q)=g^{\mu\nu}\Pi_T(s)+q^\mu q^\nu\Pi_L(s),
\label{eq:tl}
\end{equation}
where
\begin{equation}
\Pi_i(s)=\frac{G_Fz}{2^{3/2}}\left[V^2\Pi_i^V(s)+A^2\Pi_i^A(s)\right]
\qquad (i=T,L).
\label{eq:va}
\end{equation}
While, for $m=0$, only the transversal part $\Pi_T(s)$ matters, the
longitudinal part $\Pi_L(s)$, too, is relevant for $m>0$.
Due to vector current conservation, we have
\begin{equation}
\Pi_T^V(s)+s\Pi_L^V(s)=0,
\end{equation}
so that we have to consider only three different coefficient functions.
Through the two-loop order, we have
\cite{Kniehl:1989bc,Kniehl:1989yc,Kniehl:1991gu}
\begin{eqnarray}
\im\Pi_T^V(s)&=&\frac{s}{12\pi}N\left[v_0(r)+\frac{3}{4}\left(
\frac{\alpha}{\pi}Q^2+\frac{\alpha_s}{\pi}C_F\right)v_1(r)\right],
\nonumber\\
\im\Pi_T^A(s)&=&\frac{s}{12\pi}N\left[a_0(r)+\frac{3}{4}\left(
\frac{\alpha}{\pi}Q^2+\frac{\alpha_s}{\pi}C_F\right)a_1(r)\right],
\nonumber\\
\im\Pi_L^A(s)&=&\im\Pi_L^V(s)-\frac{1}{12\pi}N\left[l_0(r)+\frac{3}{4}\left(
\frac{\alpha}{\pi}Q^2+\frac{\alpha_s}{\pi}C_F\right)l_1(r)\right],
\label{eq:impi}
\end{eqnarray}
where
\begin{equation}
v_0(r)=\rho\left(1+\frac{1}{2r}\right),
\qquad
a_0(r)=\rho^3,
\qquad
l_0(r)=0,
\end{equation}
$rv_1(r)$ and $ra_1(r)$ are given by Eqs.~(5) and (6) in
Ref.~\cite{Kniehl:1989yc}, respectively, $-l_1(r)/(4\pi)$ is given by the
second term on the right-hand side of Eq.~(13) in Ref.~\cite{Kniehl:1991gu},
$r=s/(4m^2)$, and $\rho=\sqrt{1-1/r}$.
In Eq.~(\ref{eq:impi}), the normalizations are arranged so that
$v_i(\infty)=a_i(\infty)=1$ ($i=0,1$), while we have
$l_1(r)=1/r^2+{\cal O}(1/r^3)$ for $r\gg1$.

The correction terms in Eq.~(\ref{eq:dg}) may now be presented in a compact
form.
Specifically, we have
\begin{equation}
\delta_i=\frac{1}{V^2+A^2}\left\{V^2v_i(r)+A^2a_i(r)
+\frac{(s-z)^2}{C_1}\left[-\frac{\lambda^{3/2}}{3z^3}A^2
\bigl(a_i(r)-v_i(r)-l_i(r)\bigr)
+y_i\right]\right\},
\end{equation}
where $y_0$ and $y_1$ represent, respectively, the tree-level and one-loop
corrections involving one or two powers of the $Hf\overline{f}$ Yukawa
coupling.
For $m=0$, we have $\delta_i=1$, so that $\delta_i-1$ measures the relative
finite-$m$ correction at tree level ($i=0$) or at one loop in QED and QCD
($i=1$).
For $y_0$, we find
\begin{eqnarray}
y_0&=&\frac{2m^2}{z^2(z-s)}\left\{\rho\sqrt{\lambda}\left[
-\left(V^2(z+2m^2)+A^2(z-4m^2)\right)\frac{s(z-s)(h-4m^2)}{hzs+m^2\lambda}
-4V^2z+A^2
\right.\right.
\nonumber\\
&&{}\times\left.
\vphantom{\frac{s(z-s)(h-4m^2)}{hzs+m^2\lambda}}
\left(6z+s-\frac{s}{z}(2h-s)\right)\right]
+V^2L\left[3z^2-z(7h-6s)-2h^2-hs-s^2
+4m^2(5h+2s)
\vphantom{\frac{h^2-2m^2(5h-2s)+8m^4}{h+z-s}}
\right.
\nonumber\\
&&{}-\left.16m^4
+2h\frac{h^2-2m^2(5h-2s)+8m^4}{h+z-s}\right]
+A^2L\left[3z^2+z(h+6s)+(h-s)(2h+s)
\vphantom{\frac{h^2-4m^2(h+s)+16m^4}{h+z-s}}
\right.
\nonumber\\
&&{}-\left.\left.
2m^2\left(13z-h+6s+\frac{(h-s)(2h-s)}{z}\right)
+32m^4
-2h\frac{h^2-4m^2(h+s)+16m^4}{h+z-s}\right]\right\},
\nonumber\\
&&
\end{eqnarray}
where
\begin{equation}
L=\ln\frac{h+z-s-\rho\sqrt{\lambda}}{h+z-s+\rho\sqrt{\lambda}}.
\end{equation}
As we shall see in Sec.~\ref{sec:three}, the contribution to $\delta_0$
proportional to $y_0$ is exceedingly small.
Our result for $y_1$ is too lengthy to be presented here.

Furthermore, we have
\begin{equation}
\delta_\text{w}=2\re\Delta_\text{weak}+\delta_m,
\end{equation}
where the $m=0$ part $\Delta_\text{weak}$ is listed in analytic form in
Ref.~\cite{Kniehl:1991hk} and $\delta_m$ comprises the finite-$m$ correction.
We include $\delta_m$ in our numerical analysis, although it turns out to be
small against $\Delta_\text{weak}$, but we refrain from presenting here our
analytic expression for it because it is too lengthy.

Finally, using the improved Born approximation (IBA) \cite{Consoli:1989pc}, we
obtain
\begin{eqnarray}
\delta_t&=&\frac{(1+\delta_{ZZH})^2}{1-\Delta\rho}\,
\frac{(V-4c_w^2Q\Delta\rho)^2+A^2}{V^2+A^2}
\nonumber\\
&=&1+2\delta_{ZZH}+(1-8X)\Delta\rho
+\delta_{ZZH}^2+2(1-8X)\delta_{ZZH}\Delta\rho+(1-8X+16Y)(\Delta\rho)^2
\nonumber\\
&&{}+{\cal O}(x_t^3),
\label{eq:ho}
\end{eqnarray}
where $(1+\delta_{ZZH})$ is the correction to the heavy-top-quark effective
Lagrangian of the $ZZH$ interaction, $\Delta\rho=1-1/\rho$ measures the
deviation of the electroweak $\rho$ parameter from unity,
$X=c_w^2QV/(V^2+A^2)$, and $Y=c_w^4Q^2/(V^2+A^2)$.
Including the corrections of
${\cal O}(x_t)$
\cite{Fleischer:1980ub,Kniehl:1990mq,Bardin:1991dp,Dawson:1988th,Veltman:1977kh},
${\cal O}(x_t^2)$ \cite{Djouadi:1997rj,vanderBij:1986hy},
${\cal O}(x_t\alpha_s)$ \cite{Kniehl:1995gj,Djouadi:1987gn},
and ${\cal O}(x_t\alpha_s^2)$ \cite{Kniehl:1995at,Avdeev:1994db},
we have
\begin{eqnarray}
\delta_{ZZH}&=&x_t\left\{-\frac{5}{2}-\left[\frac{177}{8}+18\zeta(2)\right]x_t
+[15-2\zeta(2)]a_s+17.117\,a_s^2\right\},
\nonumber\\
\Delta\rho&=&x_t\left\{3+3\left[19-12\zeta(2)\right]x_t
-2[1+2\zeta(2)]a_s-43.782\,a_s^2\right\},
\label{eq:dr}
\end{eqnarray}
where $\zeta(2)=\pi^2/6$ and $a_s=\alpha_s^{(6)}(m_t)/\pi$.
Analytic expressions for the ${\cal O}(x_t\alpha_s^2)$ terms in
Eq.~(\ref{eq:dr}) may be found in Refs.~\cite{Kniehl:1995at,Avdeev:1994db}.
To avoid double counting, the ${\cal O}(x_t)$ term of Eq.~(\ref{eq:ho})
\cite{Kniehl:1991hk,Kniehl:1992qq},
\begin{equation}
\delta_{x_t}=-2x_t(1+12X),
\end{equation}
is subtracted in Eq.~(\ref{eq:dg}).

A more conservative form of Eq.~(\ref{eq:dg}) is obtained by discarding the
quantities $\delta_{x_t}$ and $\delta_t$, and in turn introducing within the
square brackets the term $\delta_\text{ho}$ that contains the leading
top-quark-induced corrections beyond one loop.
In the leptonic case $f=l$, which is of special interest here, we have
\begin{eqnarray}
\delta_\text{ho}&=&\delta_t-1-\delta_{x_t}
\nonumber\\
&=&\{13-72\zeta(2)+24[-17+12\zeta(2)]X+144Y\}x_t^2
\nonumber\\
&&{}+4\{7-2\zeta(2)+4[1+2\zeta(2)]X\}x_ta_s
+(-9.548+350.257\,X)x_ta_s^2.
\end{eqnarray}

Finally, $\Gamma_{H\to Zf\overline{f}}$ is obtained by integrating
Eq.~(\ref{eq:dg}) over the interval $4m^2<s<(m_H-m_Z)^2$.
The tree-level result for $m=0$ reads \cite{Keung:1984hn}
\begin{equation}
\Gamma_{H\to Zf\overline{f}}^0=\frac{G_F^2z^4}{128\pi^3h^{3/2}}
N(V^2+A^2)\left[-F\left(\frac{h}{z}\right)\right],
\label{eq:g0}
\end{equation}
where $F(x)$ is given in Eq.~(3) of Ref.~\cite{Kniehl:1992qq}.\footnote{%
In the journal publication \cite{Kniehl:1992qq}, this equation contains a
misprint, which is absent in the preprint:
$\sqrt{\frac{1}{4}x-x}$ should replaced by $\sqrt{\frac{x}{4-x}}$.}
For the reader's convenience, we reproduce this formula here:
\begin{eqnarray}
F(x)&=&\frac{1-x}{x}\left(-\frac{47}{2}+\frac{13}{2}x-x^2\right)
+\left(-2+3x-\frac{x^2}{2}\right)\ln x
\nonumber\\
&&{}+\left(10-4x+\frac{x^2}{2}\right)\sqrt{\frac{x}{4-x}}
\left(\pi-6\arcsin\frac{\sqrt{x}}{2}\right).
\end{eqnarray}
The origin of the minus sign on the right-hand side of Eq.~(\ref{eq:g0}) is
explained in Eq.~(\ref{eq:rel}).

\boldmath
\subsection{$H\to W^+f\overline{f^\prime}$ decay}
\label{sec:twob}
\unboldmath

We now consider the decay process $H\to W^+f\overline{f^\prime}$, where $f$ is
a generic fermion with weak isospin $I=-1/2$ and $f^\prime$ is an appropriate
fermion with $I=1/2$.
By charge-conjugation invariance, the process $H\to W^-f^\prime\overline{f}$
has the same partial decay width.

Some of the expressions for $H\to Zf\overline{f}$ in Sec.~\ref{sec:twoa} carry
over to $H\to W^+f\overline{f^\prime}$.
Specifically, the counterparts of Eqs.~(\ref{eq:rel}), (\ref{eq:dg0}),
(\ref{eq:tl}), (\ref{eq:va}), and (\ref{eq:g0}) are obtained by substituting
$z\to w$ and $V,A\to\sqrt{2}$, and including the overall factor
$|V_{f^\prime f}|^2$, where $V_{f^\prime f}$ is the Cabibbo-Kobayashi-Maskawa
quark mixing matrix, if $f$ and $f^\prime$ are quarks.
However, structural differences occur because the intermediate boson is now
electrically charged.
In particular, the separation of QED and weak corrections is no longer
meaningful at one loop because the photonic loop diagrams, including the
appropriate counterterm diagrams, no longer form a gauge-independent and
ultraviolet-finite subset \cite{Kniehl:1993ay,Kniehl:1991xe}.
Furthermore, in the case when $f$ and $f^\prime$ are quarks, the one-loop QCD
correction is no longer proportional to the QED correction.
As a consequence, the distribution of the partial decay width
$\Gamma_{H\to W^+f\overline{f^\prime}}$ in the $f\overline{f^\prime}$ invariant
mass square $s=q^2$ now takes the form
\begin{equation}
\frac{d\Gamma_{H\to W^+f\overline{f^\prime}}}{ds}=
\frac{d\Gamma_{H\to W^+f\overline{f^\prime}}^0}{ds}\left(\delta_0
+\frac{3}{4}\,\frac{\alpha_s}{\pi}C_F\,\delta_1
+\delta_\text{ew}-\delta_{x_t}+\delta_\text{res}\right)\delta_t.
\label{eq:dgw}
\end{equation}

In view of $m_\nu\ll m_\ell$ and $m_s\ll m_c$, we may safely neglect the mass
of the lighter one of the two fermions $f$ and $f^\prime$ and call the mass of
the heavier one $m$.
Due to the $\gamma_5$ reflection property
$\Pi_i^V(s,m_1,m_2)=\Pi_i^A(s,m_1,-m_2)$ ($i=T,L$), we then have
\cite{Kniehl:1989bc,Kniehl:1989yc,Kniehl:1991gu}
\begin{equation}
\Pi_i^V(s)=\Pi_i^A(s).
\end{equation}
Through the two-loop order, we have
\cite{Kniehl:1989bc,Kniehl:1989yc,Kniehl:1991gu}
\begin{eqnarray}
\im\Pi_T^V(s)&=&\frac{s}{12\pi}N\left[f_0(x)
+\frac{3}{4}\,\frac{\alpha_s}{\pi}C_Ff_1(x)\right],
\nonumber\\
\im\Pi_L^V(s)&=&-\frac{1}{12\pi}N\left[g_0(x)
+\frac{3}{4}\,\frac{\alpha_s}{\pi}C_Fg_1(x)\right],
\end{eqnarray}
where
\begin{equation}
f_0(x)=\left(1+\frac{1}{2x}\right)\left(1-\frac{1}{x}\right)^2,
\qquad
g_0(x)=\left(1+\frac{2}{x}\right)\left(1-\frac{1}{x}\right)^2,
\end{equation}
$xf_1(x)/4$ is given by Eq.~(7) in Ref.~\cite{Kniehl:1989yc}, $-g_1(x)/4$ is
given by Eq.~(14) in Ref.~\cite{Kniehl:1991gu}, and $x=s/m^2$.
In Eq.~(\ref{eq:impi}), the normalizations are arranged so that
$f_i(\infty)=g_i(\infty)=1$ ($i=0,1$).

Then, we have
\begin{equation}
\delta_i=f_i(x)+\frac{(s-w)^2}{C_1}
\left\{-\frac{\lambda^{3/2}}{3w^3}[f_i(x)-g_i(x)]+y_i\right\},
\end{equation}
where $C_1$ and $\lambda$ are defined as below Eq.~(\ref{eq:dg0}), but with $z$
replaced by $w$.
For $y_0$, we find
\begin{eqnarray}
y_0&=&\frac{m^2}{2w^3s^2(w-s)}\left\{
\sqrt{\lambda}\left[
\vphantom{\frac{m^2s^2(w-s)(2w+m^2)(w+s-2m^2)^2}{h(w-m^2)(s-m^2)+m^2(w-s)^2}}
s^3(3w+s)
-2h(s-m^2)\bigl(s^2+2m^2(w-s)\bigr)+2m^2w(w+s)
\right.\right.
\nonumber\\
&&{}\times\left.(s-2m^2)
+\frac{m^2s^2(w-s)(2w+m^2)(w+s-2m^2)^2}{h(w-m^2)(s-m^2)+m^2(w-s)^2}
\right]
+s^2L\left[2w\bigl(3w(-h+w+2s)
\right.
\nonumber\\
&&{}-\left.
\vphantom{\frac{m^2s^2(w-s)(2w+m^2)(w+s-2m^2)^2}{h(w-m^2)(s-m^2)+m^2(w-s)^2}}
\left.
s(h+s)\bigr)+m^2\bigl(w(9h-5w-2s)+s(3h-s)\bigr)
-4m^4(h+w+s)\right]\right\},
\end{eqnarray}
where
\begin{equation}
L=\ln\frac{(s+m^2)(w-s)+(s-m^2)(h-\sqrt{\lambda})}
{(s+m^2)(w-s)+(s-m^2)(h+\sqrt{\lambda})}.
\end{equation}
As will be shown in Sec.~\ref{sec:three}, the contribution proportional to
$y_0$ is very small compared to $\delta_0$.
For lack of space, we do not present $y_1$ and $\delta_\text{ew}$ in analytic
form.
Furthermore, we have
\begin{equation}
\delta_t=(1+\delta_{WWH})^2.
\label{eq:wwh}
\end{equation}
Including the corrections of
${\cal O}(x_t)$
\cite{Fleischer:1980ub,Kniehl:1991xe,Bardin:1991dp,Dawson:1988th},
${\cal O}(x_t^2)$ \cite{Djouadi:1997rj},
${\cal O}(x_t\alpha_s)$ \cite{Kniehl:1995ra},
and ${\cal O}(x_t\alpha_s^2)$ \cite{Kniehl:1995at},
we have
\begin{equation}
\delta_{WWH}=x_t\left\{-\frac{5}{2}+\left[\frac{39}{8}-18\zeta(2)\right]x_t
+[9-2\zeta(2)]a_s+27.041\,a_s^2\right\}.
\end{equation}
An analytic expression for the ${\cal O}(x_t\alpha_s^2)$ term may be found in
Ref.~\cite{Kniehl:1995at}.
The ${\cal O}(x_t)$ term of Eq.~(\ref{eq:wwh}) is
\begin{equation}
\delta_{x_t}=-5x_t.
\end{equation}
Again, we may trade the quantities $\delta_{x_t}$ and $\delta_t$ in
Eq.~(\ref{eq:dgw}) against the term $\delta_\text{ho}$, carrying the leading
top-quark-induced corrections beyond one loop, to be inserted within the
parentheses.
In the leptonic case, we have
\begin{eqnarray}
\delta_\text{ho}&=&\delta_t-1-\delta_{x_t}
\nonumber\\
&=&4[4-9\zeta(2)]x_t^2+2[9-2\zeta(2)]x_ta_s+54.082\,x_ta_s^2.
\end{eqnarray}

Finally, $\Gamma_{H\to W^+f\overline{f^\prime}}$ is obtained by integrating
Eq.~(\ref{eq:dgw}) over the interval $m^2<s<(m_H-m_W)^2$.

\section{Numerical results}
\label{sec:three}

We are now in a position to explore the phenomenological consequences of our
results by performing a numerical analysis.
We adopt all the input parameter values from Ref.~\cite{Nakamura:2010zzi}.
Specifically, we use
$m_W = 80.385$~GeV,
$m_Z = 91.1876$~GeV,
$m_b = 4.78$~GeV,
$m_t = 173.5$~GeV,
and $\alpha_s^{(6)}(m_t)=0.1080$,
which follows from $\alpha_s^{(5)}(m_Z)=0.1184$ via four-loop evolution and
three-loop matching \cite{Chetyrkin:1997sg}.
In our renormalization scheme,
$\sin^2\theta_\text{w}=1-m_W^2/m_Z^2=0.222897$ and
$\alpha=\sqrt{2}G_F\sin^2\theta_\text{w}m_W^2/\pi=1/132.233$ are derived
parameters.

\begin{figure}
\begin{center}
\begin{tabular}{ccc}
\includegraphics[width=0.3\textwidth]{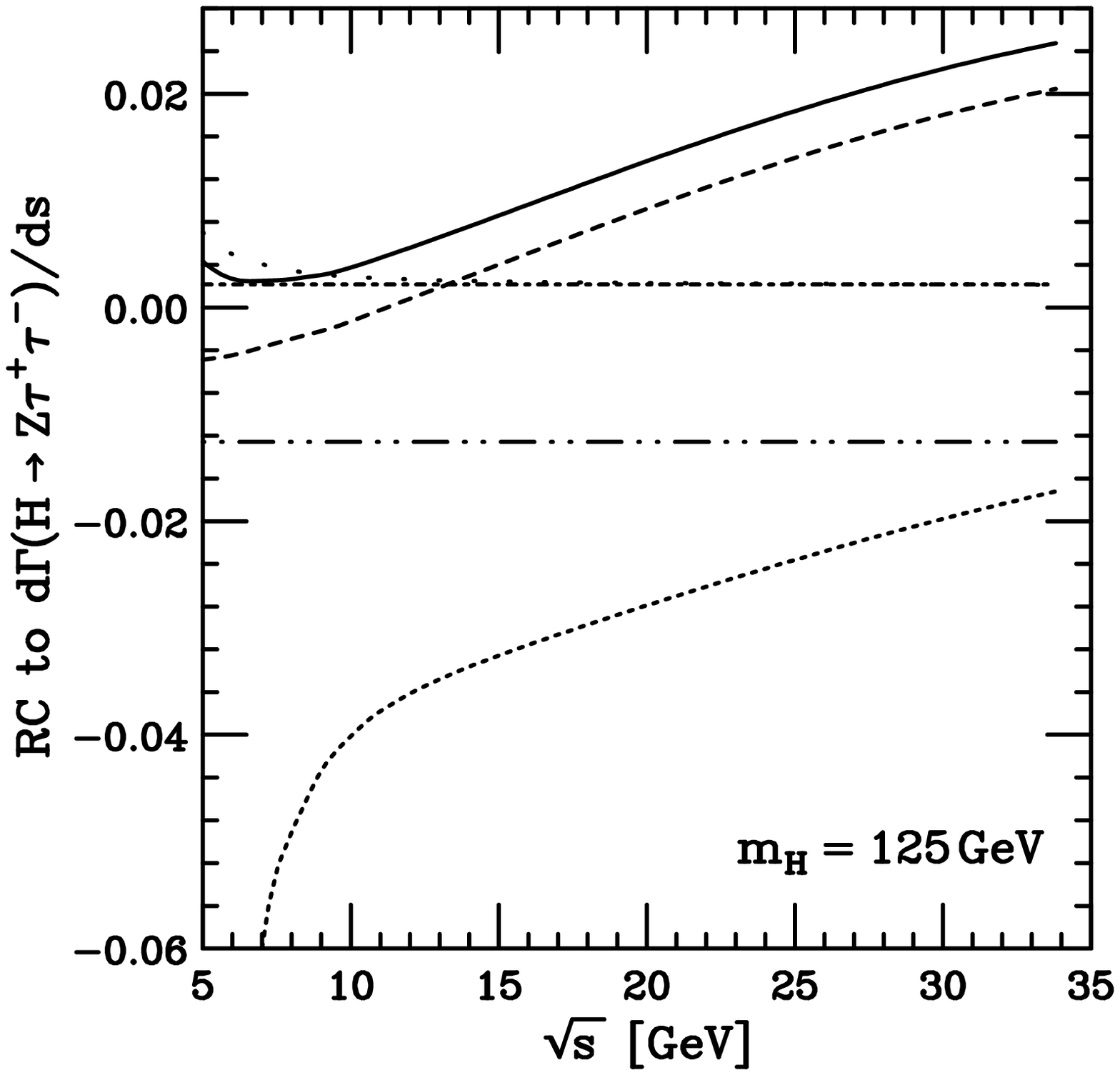}
&
\includegraphics[width=0.3\textwidth]{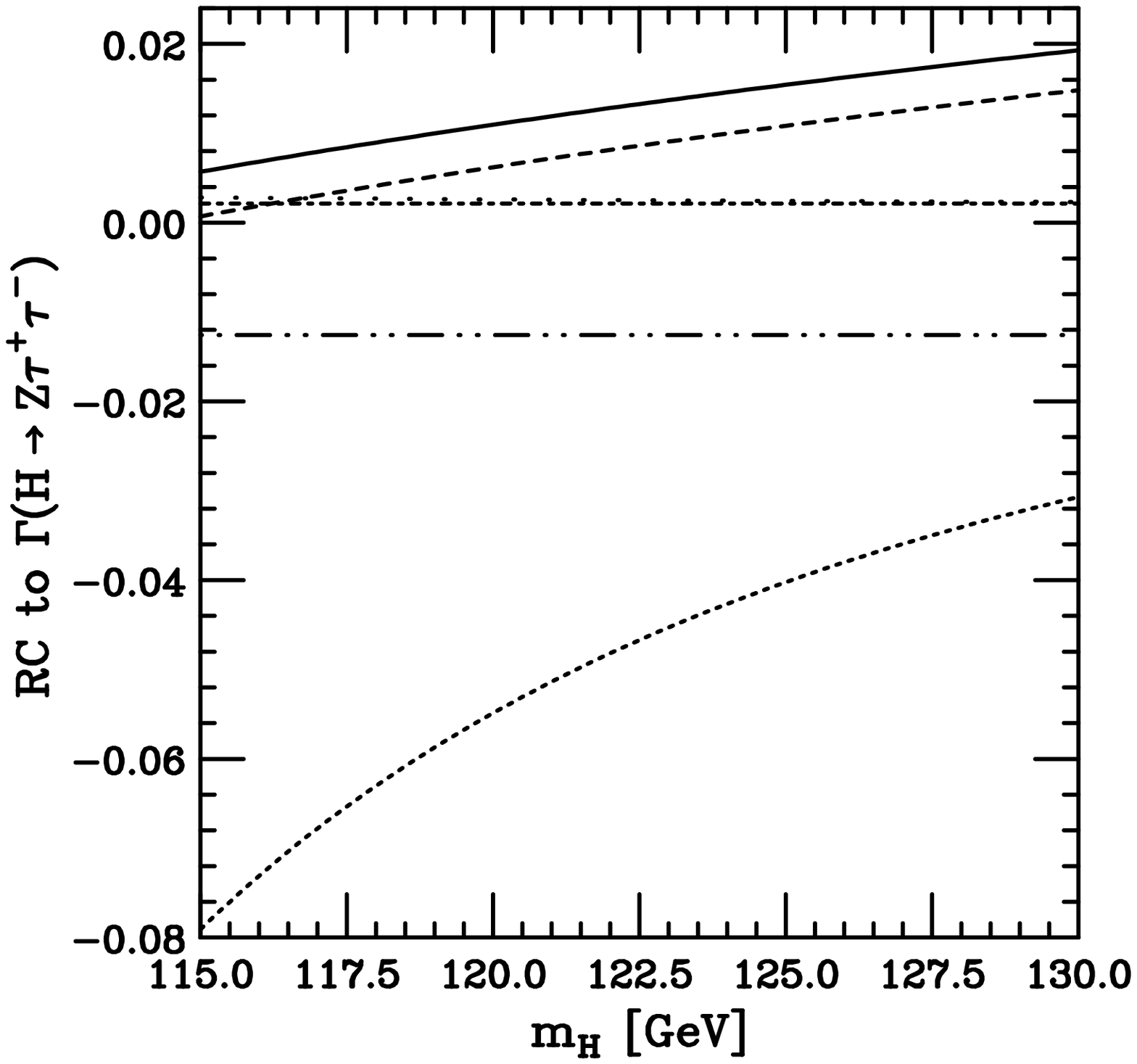}
&
\includegraphics[width=0.3\textwidth]{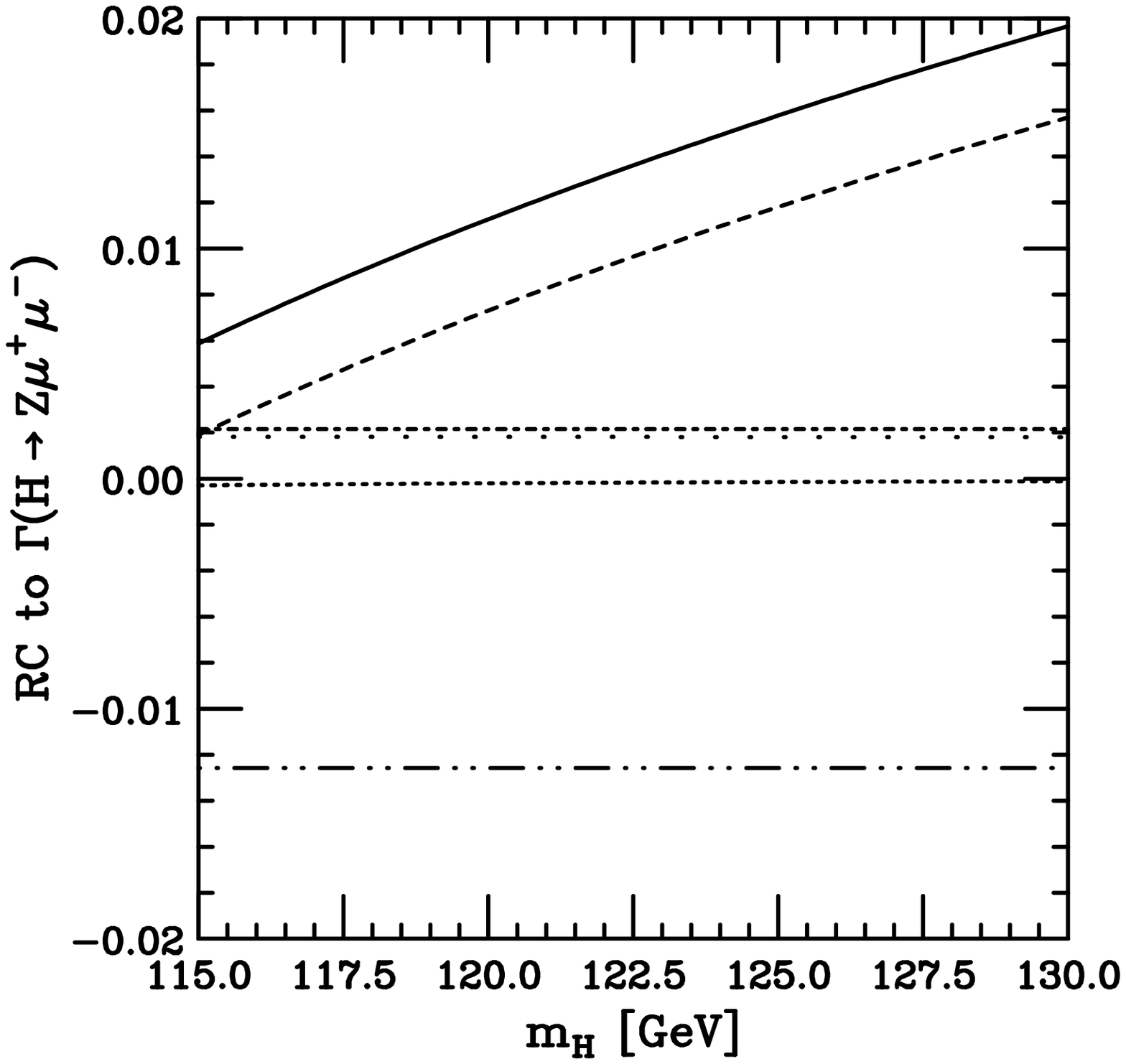}
\\
(a) & (b) & (c)
\end{tabular}
\end{center}
\caption{\label{fig:rczll}%
Tree-level finite-$m$ (dotted lines) and radiative (solid lines) corrections
to (a) $d\Gamma(H\to Z\tau^+\tau^-)/ds$ for $m_H=125$~GeV as functions of the
$\tau^+\tau^-$ invariant mass $\sqrt{s}$,
and to (b) $\Gamma(H\to Z\tau^+\tau^-)$
and (c) $\Gamma(H\to Z\mu^+\mu^-)$ as functions of $m_H$.
The radiative corrections include the
${\cal O}(\alpha)$ QED (coarsely dotted lines),
${\cal O}(\alpha)$ weak (dashed lines),
and dominant higher-order (dot-dashed lines) corrections of ${\cal O}(x_t^2)$,
${\cal O}(x_t\alpha_s)$, and ${\cal O}(x_t\alpha_s^2)$.
For comparison, the ${\cal O}(\alpha)$ corrections predicted by the IBA 
(dot-dot-dashed lines) are also shown.}
\end{figure}
We first consider the $H\to Z\ell^+\ell^-$ decays. 
In Fig.~\ref{fig:rczll}, we present the tree-level finite-$m$ (dotted lines)
and radiative (solid lines) corrections to (a) $d\Gamma(H\to Z\tau^+\tau^-)/ds$
for $m_H=125$~GeV as functions of $\sqrt{s}$, and to (b)
$\Gamma(H\to Z\tau^+\tau^-)$ and (c) $\Gamma(H\to Z\mu^+\mu^-)$ as functions of
$m_H$.
The radiative corrections are decomposed into the ${\cal O}(\alpha)$ QED
(coarsely dotted lines), ${\cal O}(\alpha)$ weak (dashed lines), and dominant
higher-order (dot-dashed lines) corrections, of ${\cal O}(x_t^2)$,
${\cal O}(x_t\alpha_s)$, and ${\cal O}(x_t\alpha_s^2)$.
For comparison, the ${\cal O}(\alpha)$ corrections predicted by the IBA 
(dot-dot-dashed lines) are also shown.
Looking at Fig.~\ref{fig:rczll}(a), we observe that the QED correction
$3\alpha\delta_1/(4\pi)$ exhibits an enhancement towards low values of
$\sqrt{s}$, which is of Coulomb origin, while it gets close to its asymptotic
value $3\alpha/(4\pi)$ as $\sqrt{s}$ approaches its kinematic upper bound.
The purely weak correction $\delta_\text{w}$ monotonically increases with
$\sqrt{s}$, ranging from slightly negative values at the $\tau^+\tau^-$ pair
production threshold to about 2.0\% at the upper endpoint.
The one-loop electroweak correction is inadequately described by the IBA term
$\delta_{x_t}$.
The dominant higher-order correction $\delta_\text{ho}$ amounts to about 0.2\%
altogether and incidentally almost coincides with the QED correction.
The finite-$m$ correction $\delta_0-1$, of course, quenches
$d\Gamma(H\to Z\tau^+\tau^-)/ds$ at threshold, but it is still as large as
$-1.7\%$ at the upper endpoint, largely compensating the combined radiative
correction.
As anticipated in Sec.~\ref{sec:twoa}, the relative contribution of $y_0$ to
$\delta_0$, proportional to the $H\tau^+\tau^-$ coupling, is exceedingly small
in magnitude, below 0.09\%, over the full $\sqrt{s}$ range.
The finite-$m$ corrections to $d\Gamma(H\to Z\mu^+\mu^-)/ds$ and
$d\Gamma(H\to Ze^+e^-)/ds$ are negligible compared to the expected size of the
presently unknown subleading higher-order corrections $\delta_\text{res}$, and
the radiative corrections to both observables are practically indistinguishable
thanks to the almost perfect lepton universality.
The latter are also very similar to the radiative corrections to
$d\Gamma(H\to Z\tau^+\tau^-)/ds$, and we refrain from presenting the
counterparts of Fig.~\ref{fig:rczll}(a) for $d\Gamma(H\to Z\mu^+\mu^-)/ds$ and
$d\Gamma(H\to Ze^+e^-)/ds$.

Looking at Fig.~\ref{fig:rczll}(b), we observe that the finite-$m$ correction
to $\Gamma(H\to Z\tau^+\tau^-)$ ranges from $-7.9\%$ to $-3.1\%$ in the
considered mass window 115~GeV${}<m_H<130$~GeV and more than compensates the
overall radiative correction, which ranges there between 0.6\% and 1.9\%.
From Fig.~\ref{fig:rczll}(c), we read off that the finite-$m$ correction to
$\Gamma(H\to Z\mu^+\mu^-)$ is below 0.03\% in magnitude. 

\begin{figure}
\begin{center}
\begin{tabular}{cc}
\includegraphics[height=0.325\textheight]{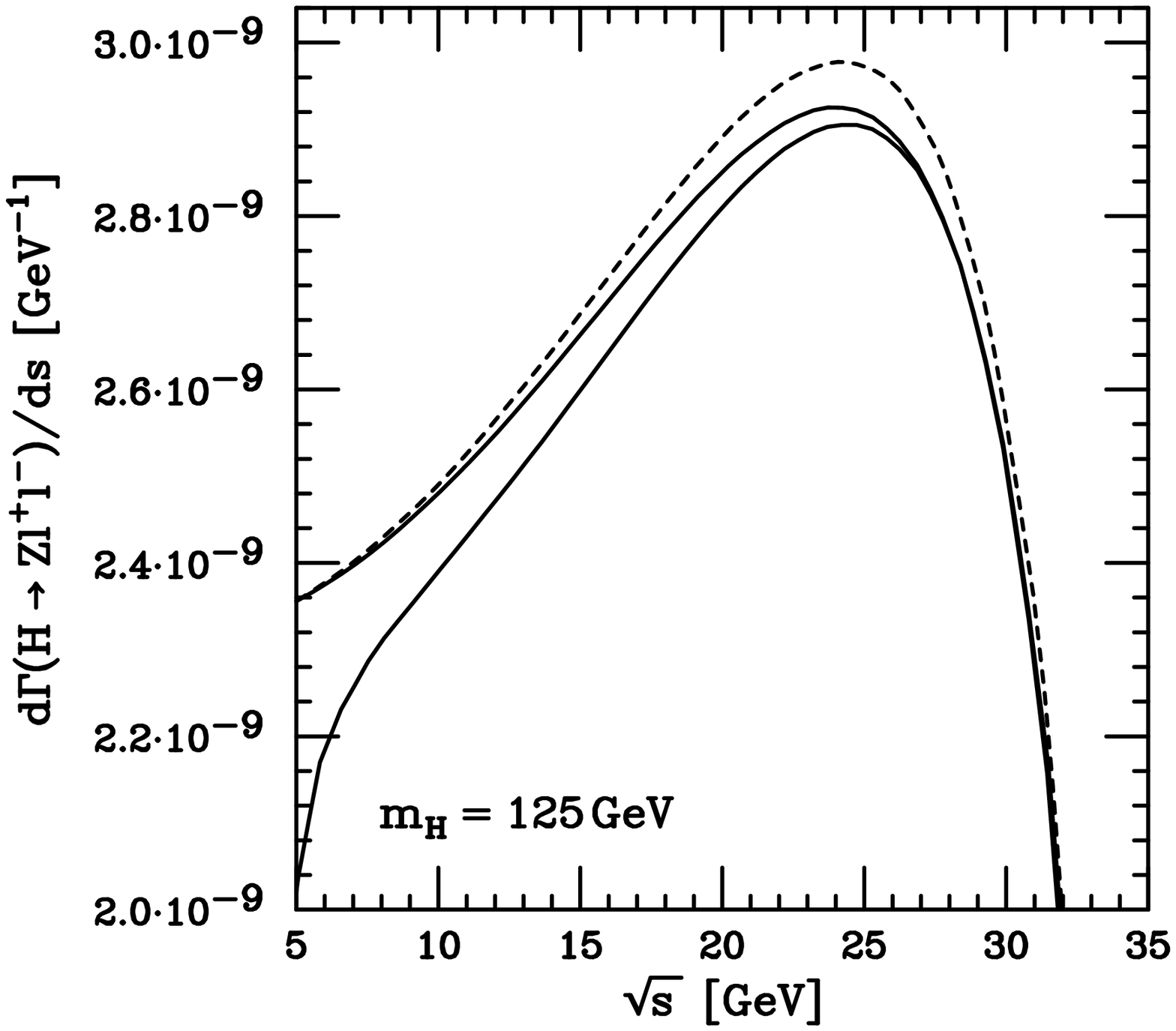}
&
\includegraphics[height=0.325\textheight]{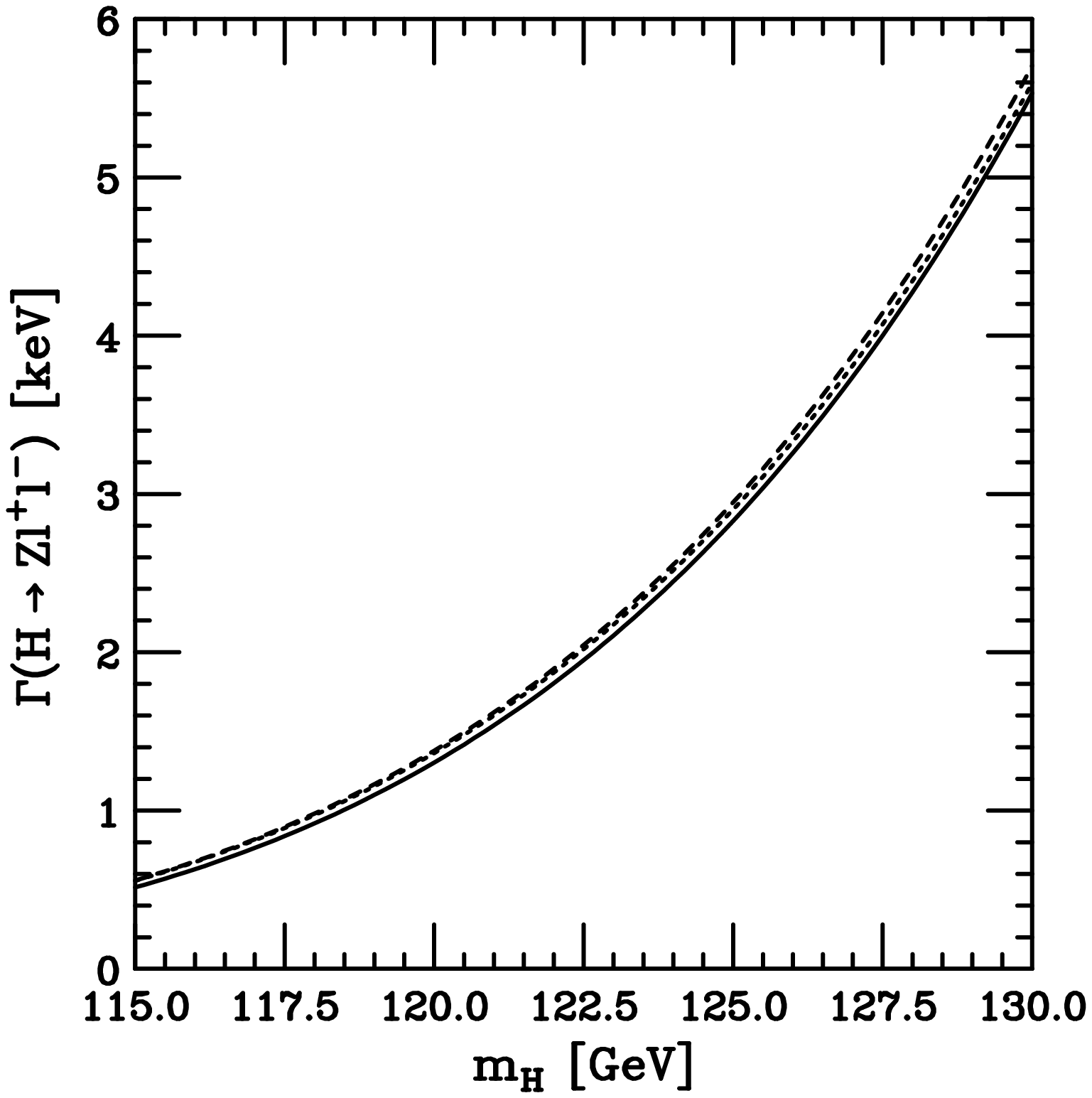}
\\
(a) & (b)
\end{tabular}
\end{center}
\caption{\label{fig:zll}%
(a) $d\Gamma(H\to Z\ell^+\ell^-)/ds$ for $m_H=125$~GeV as a function of the
$\ell^+\ell^-$ invariant mass $\sqrt{s}$, and
(b) $\Gamma(H\to Z\ell^+\ell^-)$ as a function of $m_H$
at the tree level for $m=0$ (dotted lines) and including both finite-$m$ and
radiative corrections for $\ell=\mu$ (dashed lines) and $\ell=\tau$ (solid
lines).}
\end{figure}
In Fig.~\ref{fig:zll}(a), we present our best predictions for
$d\Gamma(H\to Z\tau^+\tau^-)/ds$ (solid line) and
$d\Gamma(H\to Z\mu^+\mu^-)/ds$ (dashed line) for $m_H=125$~GeV as functions of
$\sqrt{s}$ including both finite-$m$ and radiative corrections.
For comparison, the tree-level result for $m=0$ (dotted line) is also shown.
The relation of the solid line shape to the dotted one may be easily understood
from Fig.~\ref{fig:rczll}(a).
The essential feature of the dashed line shape in comparison to the solid one
is the insignificance of the finite-$m$ correction for $\ell=\mu$ already
mentioned above.

In Fig.~\ref{fig:zll}(b), our best predictions for $\Gamma(H\to Z\tau^+\tau^-)$
(solid line) and $\Gamma(H\to Z\mu^+\mu^-)$ (dashed line) as functions of $m_H$
are compared with the tree-level result for $m=0$ (dotted line).
The relative shifts of the solid and dashed line shapes with respect to the
dotted one immediately follow from Figs.~\ref{fig:rczll}(b) and (c),
respectively.

\begin{figure}
\begin{center}
\begin{tabular}{ccc}
\includegraphics[width=0.3\textwidth]{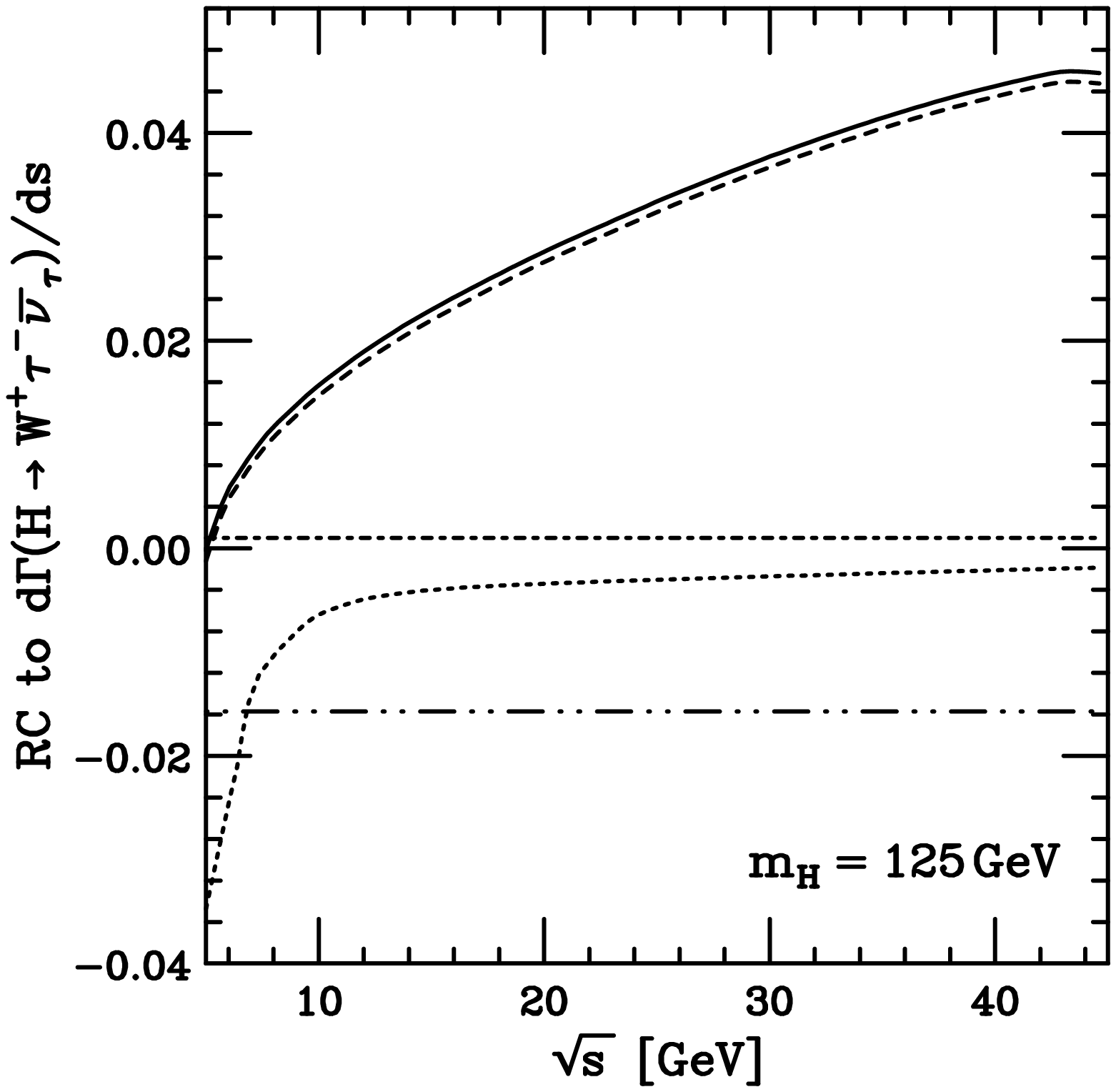}
&
\includegraphics[width=0.3\textwidth]{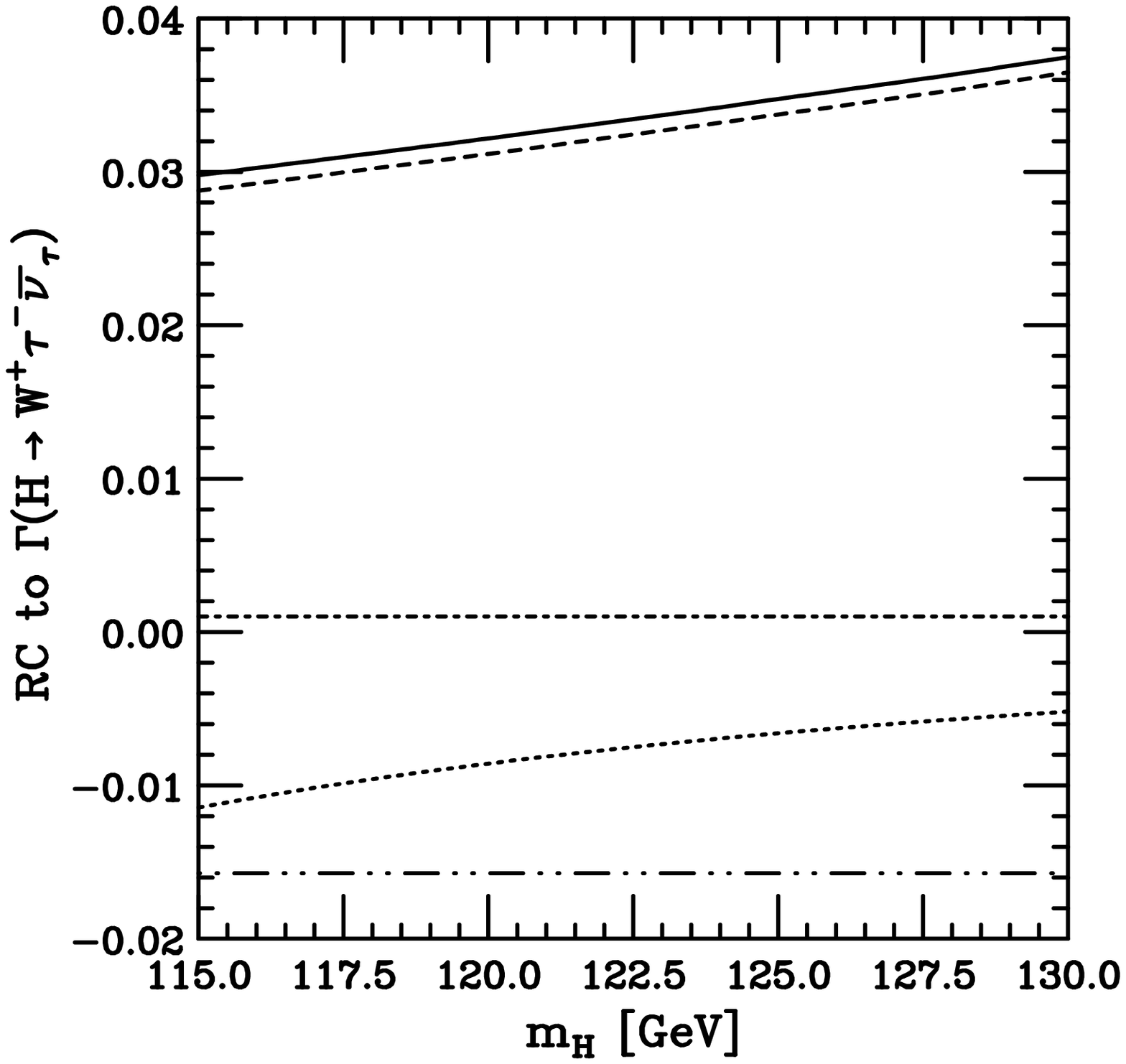}
&
\includegraphics[width=0.3\textwidth]{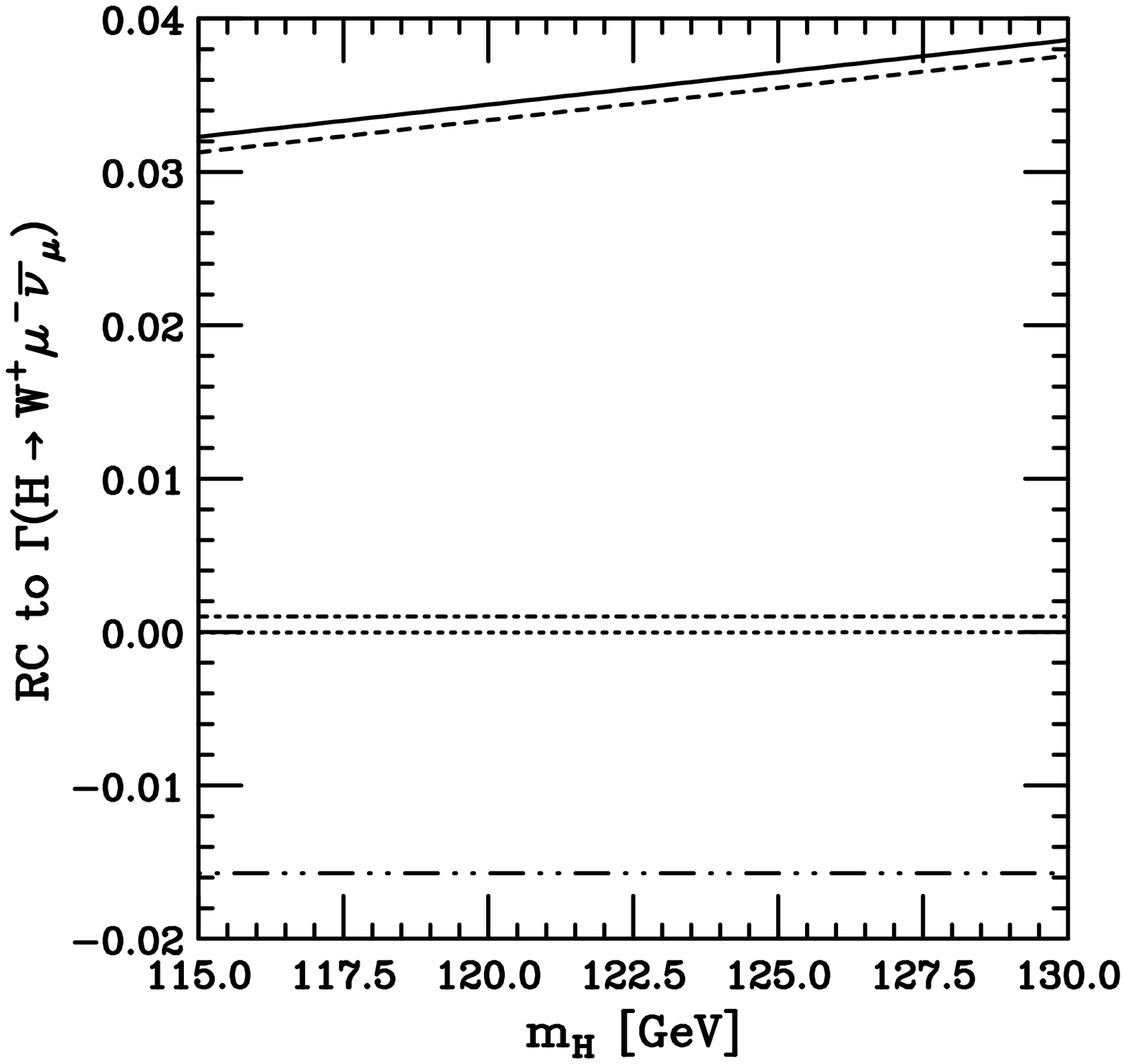}
\\
(a) & (b) & (c)
\end{tabular}
\end{center}
\caption{\label{fig:rcwln}%
Tree-level finite-$m$ (dotted lines) and radiative (solid lines) corrections
to (a) $d\Gamma(H\to W^+\tau^-\overline{\nu}_\tau)/ds$ for $m_H=125$~GeV as
functions of the $\tau^-\overline{\nu}_\tau$ invariant mass $\sqrt{s}$,
and to (b) $\Gamma(H\to W^+\tau^-\overline{\nu}_\tau)$
and (c) $\Gamma(H\to W^+\mu^-\overline{\nu}_\mu)$ as functions of $m_H$.
The radiative corrections include the
${\cal O}(\alpha)$ electroweak (dashed lines)
and dominant higher-order (dot-dashed lines) corrections of ${\cal O}(x_t^2)$,
${\cal O}(x_t\alpha_s)$, and ${\cal O}(x_t\alpha_s^2)$.
For comparison, the ${\cal O}(\alpha)$ corrections predicted by the IBA 
(dot-dot-dashed lines) are also shown.}
\end{figure}
We now turn to the $H\to W^+\ell^-\overline{\nu}_\ell$ decays. 
Figure~\ref{fig:rcwln} shows the tree-level finite-$m$ (dotted lines) and
radiative (solid lines) corrections to (a)
$d\Gamma(H\to W^+\tau^-\overline{\nu}_\tau)/ds$ for $m_H=125$~GeV as functions
of $\sqrt{s}$, and to (b) $\Gamma(H\to W^+\tau^-\overline{\nu}_\tau)$ and (c)
$\Gamma(H\to W^+\mu^-\overline{\nu}_\mu)$ as functions of $m_H$.
The radiative corrections are built up by the ${\cal O}(\alpha)$ electroweak
(dashed lines) and dominant higher-order (dot-dashed lines) corrections of
${\cal O}(x_t^2)$, ${\cal O}(x_t\alpha_s)$, and ${\cal O}(x_t\alpha_s^2)$.
For comparison, the ${\cal O}(\alpha)$ corrections predicted by the IBA 
(dot-dot-dashed lines) are also presented.
Looking at Fig.~\ref{fig:rcwln}(a), we observe that the electroweak correction
$\delta_\text{ew}$ monotonically increases with $\sqrt{s}$, ranging from
slightly negative values at the $\tau^-\overline{\nu}_\tau$ pair production
threshold to about 4.5\% at the upper endpoint.
The one-loop electroweak correction is again inadequately described by the
IBA term $\delta_{x_t}$.
The dominant higher-order correction $\delta_\text{ho}$ amounts to about 0.1\%
altogether.

Comparing Figs.~\ref{fig:rcwln}(a)--(c) with Figs.~\ref{fig:rczll}(a)--(c), we
observe that the finite-$m$ corrections for the
$H\to W^+\ell^-\overline{\nu}_\ell$ decays are significantly smaller than for
the respective $H\to Z\ell^+\ell^-$ decays.
This is mainly due to the fact that the shrinkage of the available phase space
caused by switching on the finite-$m$ corrections is lesser for the
$H\to W^+\ell^-\overline{\nu}_\ell$ decays because the lepton pair production
threshold is twice as low, at $\sqrt{s}=m$, and the upper endpoint is located
higher, by $m_Z-m_W\approx11$~GeV, also leading to correspondingly higher
maxima of the $\sqrt{s}$ distributions.
In fact, the finite-$m$ correction $\delta_0-1$ to
$d\Gamma(H\to W^+\tau^-\overline{\nu}_\tau)/ds$ in Fig.~\ref{fig:rcwln}(a)
rapidly relaxes from its threshold value of $-100\%$ to moderate values,
passing $-1\%$ at $\sqrt{s}\approx8$~GeV and reaching $-0.2\%$ at the upper
endpoint.  
As anticipated in Sec.~\ref{sec:twob}, the relative contribution of $y_0$ to
$\delta_0$, proportional to the $H\tau^+\tau^-$ coupling, is very small, below
0.17\%, over the full $\sqrt{s}$ range.

From Fig.~\ref{fig:rcwln}(b), we learn that the finite-$m$ correction to
$\Gamma(H\to W^+\tau^-\overline{\nu}_\tau)$ ranges from $-1.1\%$ to $-0.5\%$ in
the considered mass window 115~GeV${}<m_H<130$~GeV and reduces the radiative
corrections by between 38\% to 14\%.
On the other hand, Fig.~\ref{fig:rcwln}(c) tells us that the finite-$m$
correction to $\Gamma(H\to W^+\mu^-\overline{\nu}_\mu)$ is absolutely
negligible, being below 0.004\% in magnitude.

\begin{figure}
\begin{center}
\begin{tabular}{cc}
\includegraphics[height=0.325\textheight]{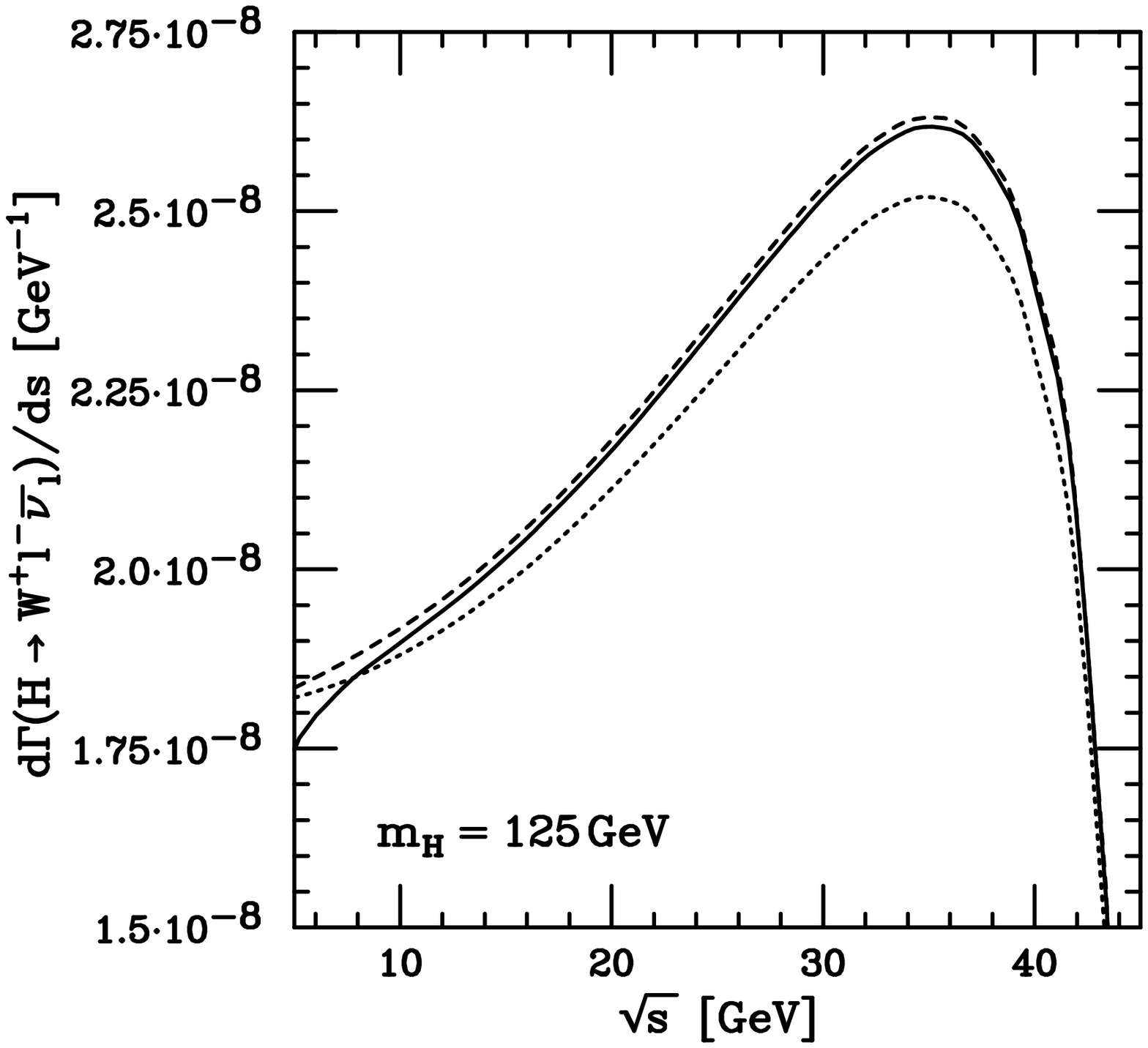}
&
\includegraphics[height=0.325\textheight]{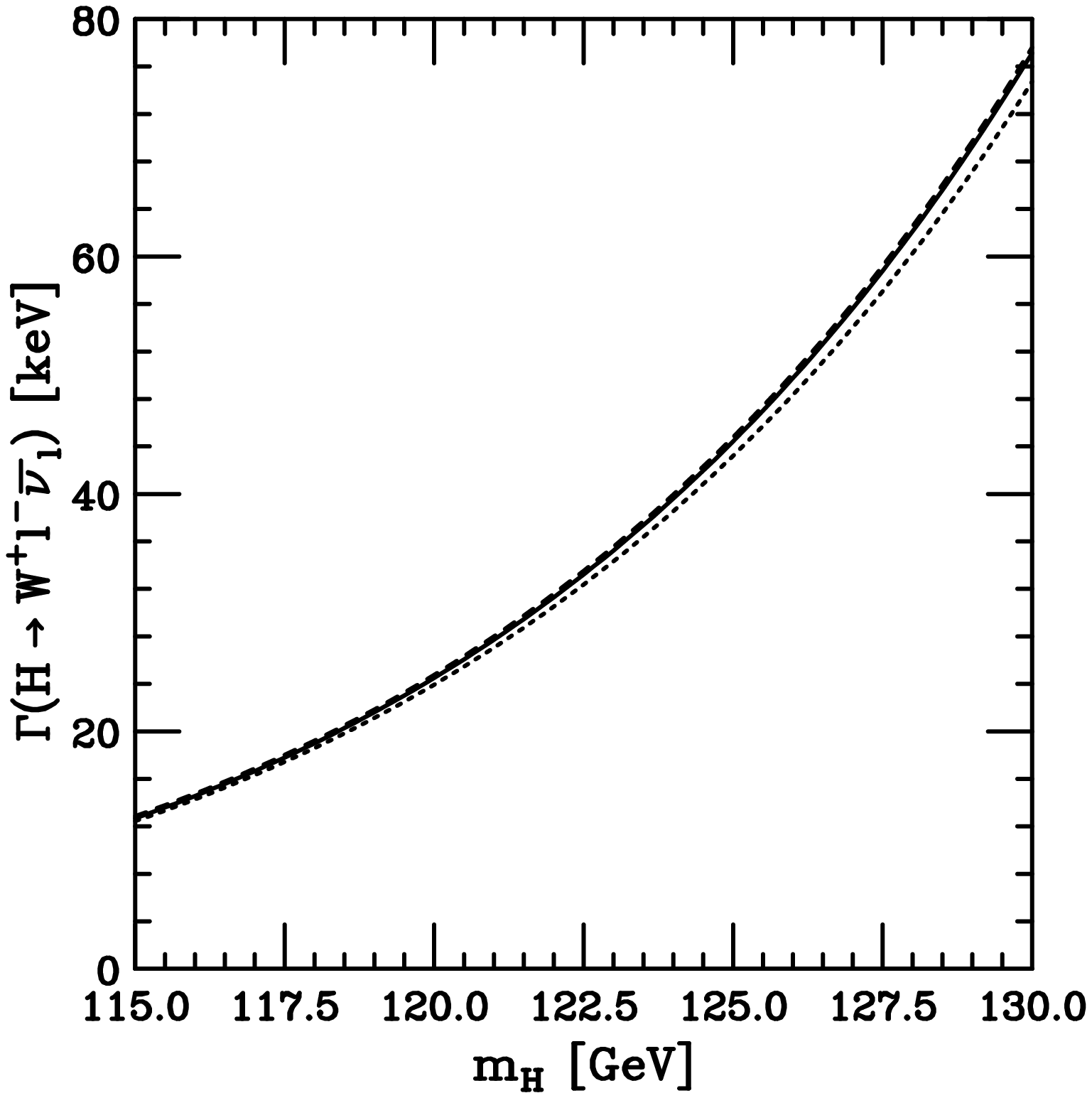}
\\
(a) & (b)
\end{tabular}
\end{center}
\caption{\label{fig:wln}%
(a) $d\Gamma(H\to W^+\ell^-\overline{\nu}_\ell)/ds$ for $m_H=125$~GeV as a
function of the $\ell^-\overline{\nu}_\ell$ invariant mass $\sqrt{s}$, and
(b) $\Gamma(H\to W^+\ell^-\overline{\nu}_\ell)$ as a function of $m_H$
at the tree level for $m=0$ (dotted lines) and including both finite-$m$ and
radiative corrections for $\ell=\mu$ (dashed lines) and $\ell=\tau$ (solid
lines).}
\end{figure}
In Fig.~\ref{fig:wln}(a), our final predictions for
$d\Gamma(H\to W^+\tau^-\overline{\nu}_\tau)/ds$ (solid line) and
$d\Gamma(H\to W^+\mu^-\overline{\nu}_\mu)/ds$ (dashed line) for $m_H=125$~GeV
are shown as functions of $\sqrt{s}$ and compared with the tree-level result
for $m=0$ (dotted line), so as to expose the interplay of the finite-$m$ and
radiative corrections.
In Fig.~\ref{fig:wln}(b), the same is done for
$\Gamma(H\to W^+\tau^-\overline{\nu}_\tau)$ and
$\Gamma(H\to W^+\mu^-\overline{\nu}_\mu)$ as functions of $m_H$.

\section{Conclusions}
\label{sec:four}

We presented a precision study of the partial widths of the
$H\to Z\ell^+\ell^-$ and $H\to W^+\ell^-\overline{\nu}_\ell$ decays, including
the full one-loop electroweak corrections and the dominant contributions at two
and three loops, of ${\cal O}(x_t^2)$, ${\cal O}(x_t\alpha_s)$, and
${\cal O}(x_t\alpha_s^2)$.
We also included finite-$m$ corrections, which turned out to be indispensable
for $\ell=\tau$, but negligible for $\ell=\mu,e$.
Working in the on-mass-shell renormalization scheme with $G_F$ replacing
$\alpha$ as a basic parameter, we ensured that the radiative corrections
remained moderate in size, being devoid of large logarithms involving small
masses of charged fermions.
As for the integrated partial decay widths, we found the net corrections
relative to the tree-level results for $m=0$ \cite{Keung:1984hn} at
$m_H=125$~GeV to be
$+1.6\%$ for $H\to Ze^+e^-$ and $H\to Z\mu^+\mu^-$,
$-2.5\%$ for $H\to Z\tau^+\tau^-$,
$+3.6\%$ for $H\to W^+e^-\overline{\nu}_e$ and
$H\to W^+\mu^-\overline{\nu}_\mu$,
and $+2.8\%$ for $H\to W^+\tau^-\overline{\nu}_\tau$.

\section*{Acknowledgment}

This work was supported in part by the German Federal Ministry for Education
and Research BMBF through Grant No.\ 05H12GUE and by the Helmholtz Association
HGF through Grant No.\ Ha~101.

\end{document}